\documentclass[doublecol]{epl2} 
\usepackage{amsmath}
\usepackage{amsfonts}
\usepackage[colorlinks=true,allcolors=blue]{hyperref}
\title{Nonlinear response theory of molecular machines}
\shorttitle{Nonlinear response theory of molecular machines} 
\let\epsilon\varepsilon

\author{Michalis Chatzittofi\inst{1} \and Jaime Agudo-Canalejo\inst{1,2} \and Ramin Golestanian \inst{1,3}}
\shortauthor{Chatzittofi M. \etal}

\institute{                    
  \inst{1} Max Planck Institute for Dynamics and Self-Organization (MPI-DS), D-37077 Göttingen, Germany\\
  \inst{2} Department of Physics and Astronomy, University College London, London WC1E 6BT, United Kingdom\\
  \inst{3} Rudolf Peierls Centre for Theoretical Physics, University of Oxford, Oxford OX1 3PU, United Kingdom
}

\abstract{Chemical affinities are responsible for driving active matter systems out of equilibrium. At the nano-scale, molecular machines interact with the surrounding environment and are subjected to external forces. The mechano-chemical coupling which arises naturally in these systems reveals a complex interplay between chemical and mechanical degrees of freedom with strong impact on their active mechanism. By considering various models far from equilibrium, we show that the tuning of applied forces give rise to a nonlinear response that causes a non-monotonic behaviour in the machines' activity. Our findings have implications in understanding, designing, and triggering such processes by controlled application of external fields, including the collective dynamics of larger non-equilibrium systems where the total dissipation and performance might be affected by internal and inter-particle interactions. }

\begin{document}

\maketitle

Active matter systems \cite{Gompper2020} convert chemical energy into motion or useful work. In the biological context, the chemical energy typically comes from ATP hydrolysis \cite{borsley2022chemical}. In general, such systems may be enzymes, molecular motors \cite{julicher1997modeling,magnasco}, stochastic micro-swimmers \cite{prl2008,golestanian2009stochastic,RG2009stoch,RG2005prl}, rotors \cite{noji1997direct,junge2015atp,Yasuda1998}, and even synthetic active matter \cite{pumm2022dna,shi2024dna,klumpp2019swimming}. When modeling the motion or the activity of such active matter systems, especially in the case of self-propelled particles, the existence of an ``active force'' that creates propulsion or motion is typically assumed \cite{speck2016stochastic}.

In reality, the mechanism that generates the useful work is some non-equilibrium cyclic process that is driven by some \textit{chemical affinity} which is the free energy dissipated per cycle \cite{seifert2018stochastic}. This affinity generates an out of equilibrium current associated to the reaction velocity of the specific process, and the product of the current and the affinity is related to the total entropy production rate \cite{seifert2008stochastic}. Any external forces (conservative or otherwise) must be treated carefully due to the mechano-chemical coupling of the internal mechanism of the nano-machines with the external, spatial, or hydrodynamic degrees of freedom which determine the extraction of work or power \cite{chatzittofi2023entropy, DaddiMoussaIder2023,bebon2024thermodynamics,fritz2023thermodynamically}. 

In the past, theoretical models for how internal energy dissipation is transduced into actuation of external degrees of freedom have been used successfully to explain the enhanced diffusion \cite{illien2017diffusion}, chemotaxis \cite{agudo2018phoresis}, and synchronization \cite{agudo2021synchronization} of enzymes, the motion of rotors by using flashing ratchets \cite{pumm2022dna}, or the steps of molecular motors \cite{julicher1997modeling,lipowsky2005life}. On the other hand, these models often lack an important missing piece: how applied forces or other experimentally-controlled triggers feed back on the dynamics of the internal degrees of freedom via the mechano-chemical coupling, beyond what linear response theory can predict \cite{onsager1,degroot}.

\begin{figure*}[t]
\centering\includegraphics[width=0.99\linewidth]{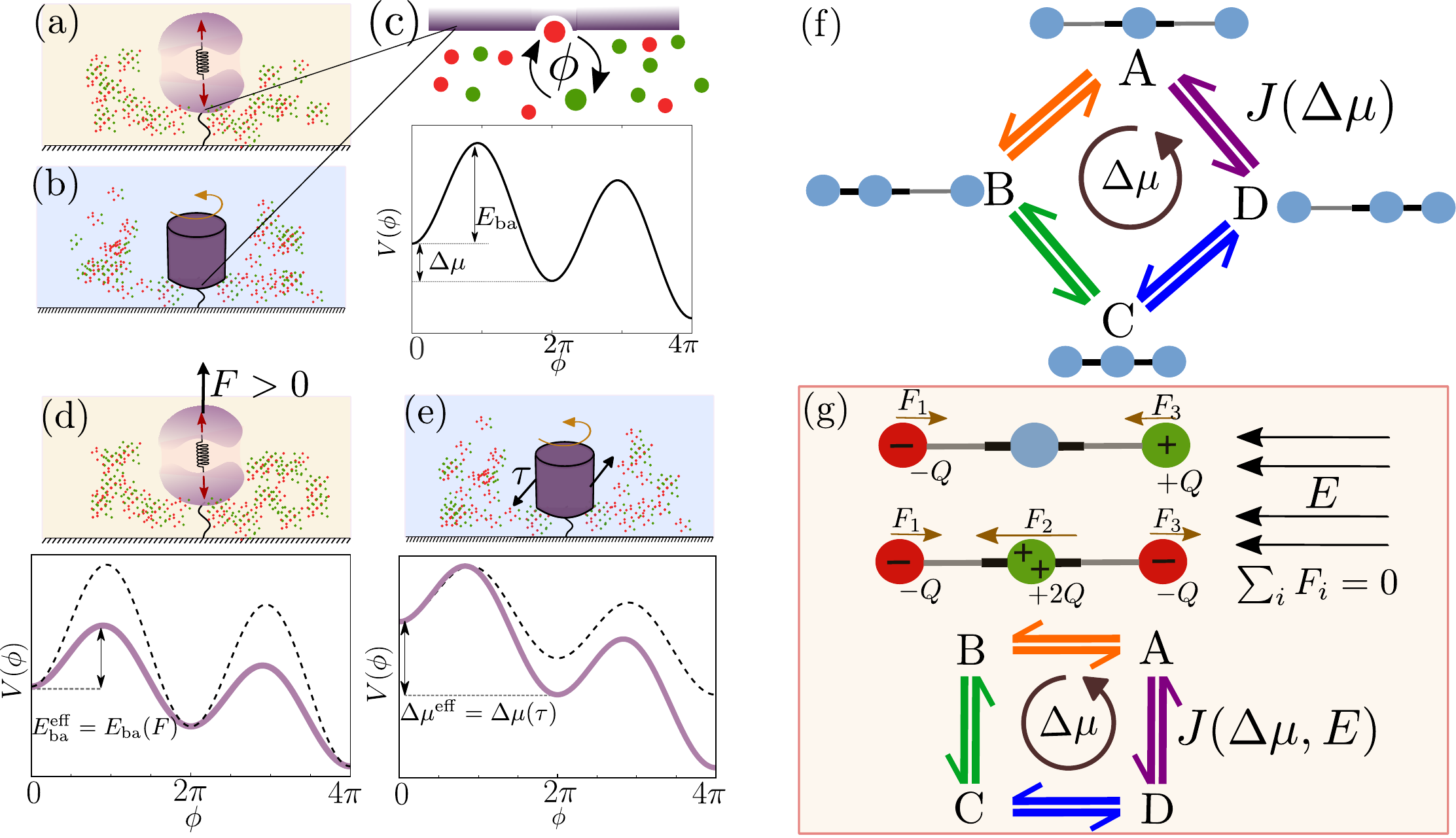}
\caption{Examples of molecular machines: (a) an enzyme that undergoes conformational changes attached to a surface, (b) a molecular rotor surrounded by a fluid medium. (c) The internal dynamics of the molecular machine involves the conversion of a `fuel' or `substrate' molecule into a lower-energy `waste' or `product' molecule. The reaction is here described by a phase $\phi$ that advances by $2\pi$ in each step along a potential $V(\phi)$ that represents the free energy landscape of the chemical reaction, with $\Delta \mu$  being the free energy difference or driving force, and $E_\mathrm{ba}$ being the energy barrier of the process. (d) For an enzyme, an applied force $F$ effectively changes the energy barrier of the internal process. (e) For a rotor, an applied torque $\tau$ effectively changes the driving force of the internal process. (f) A stochastic three-sphere swimmer model with four possible conformations \cite{prl2008}. Transitions between the states happen along a cycle driven by the affinity $\Delta \mu$, which generates a current $J(\Delta \mu)$. (g) The electric dipole $(-Q,0,+Q)$ and the electric quadrupole $(-Q,+2Q,-Q)$ models for micro-swimmers with overall charge neutrality. The arrows show the influence of the external electric field $E$ that generates electrostatic forces with vanishing net sum. The cycle affinity is still equal to $\Delta \mu$, but the current now depends also on the strength of the electric field $E$.}
	\label{fig:figure1}
\end{figure*}

In this Letter, we focus on the activity of enzymes (see Fig.~\ref{fig:figure1}(a)), rotors (Fig.~\ref{fig:figure1}(b)), and force-free micro-swimmers (Fig.~\ref{fig:figure1}(f)) in the presence of external forces. The key ingredient for activity is a chemical affinity $\Delta \mu$ which drives the system out of equilibrium and induces motion or performs work, depending on the system. To model discrete chemical reactions that dissipate an energy $\Delta \mu$, we assume that nano-machines carry an internal reaction coordinate $\phi$ which is subject to a washboard potential, see Fig.~\ref{fig:figure1}(c), where a reaction corresponds to the phase $\phi$ advancing by $2\pi$.
We show that an applied force effectively modifies either the energy barriers of the energy free landscape that governs the dynamics of the internal process or reaction coordinate (Fig.~\ref{fig:figure1}(d)), the chemical driving force (Fig.~\ref{fig:figure1}(e)), or both. This can be used to tune the catalytic activity of enzymes and motors in both directions, from enhancement to stalling. Experimentally, such manipulations might arise from ultrasound irradiation, electromagnetic fields, and other methods \cite{dik2023propelling,gumpp2009triggering}. Various experiments have reported a non-monotonic behaviour of the enzymatic activity due to such external manipulations \cite{dik2023propelling}.

In the case of micro-swimmers, to emphasize how a mechano-chemical coupling cannot generally be modelled as an active force, we consider examples of neutrally charged swimmers with spatially nonuniform charge distributions subjected to an external electric field (Fig.~\ref{fig:figure1} (g)) which applies different forces on the different sub-units of the complex. These serve as examples where applied forces do not cause any net drift or direct hydrodynamic dissipation. In particular, this simple consideration highlights how, for force-free micro-swimmers, external fields affect the activity and the total chemical energy dissipation in a nontrivial way.

\section{Enzymes}
\subsection{Model}
To model an enzyme that can undergo conformational changes, we consider a minimal model in which the enzyme is represented with two sub-units reflecting its softest mode, as shown in Fig.~\ref{fig:figure1}(a) \cite{illien2017diffusion}. We denote with $x_1$ and $x_2$ the positions of the two sub-units. To model its catalytic cycle (e.g.~ATP hydrolysis in the biological case), we use a cyclic coordinate $\phi$, with $\Delta \mu$ being the difference in Gibbs free energy catalyzed in every reaction; see Fig.~\ref{fig:figure1}(c). The non-equilibrium potential $U(x_1, x_2, \phi)$, which governs the coupled dynamics of the two spatial coordinates and the internal coordinate, reads \cite{agudo2021synchronization}
\begin{equation}\label{eq:potentialU}
    U(x_1,x_2,\phi) = 
    \tilde{V}(\phi) - f\phi+
    \frac{k}{2}\big(x_1 - x_2 - L(\phi)\big)^2 - F x_1.
\end{equation}
Here, $\tilde{V}(\phi)$ is a $2\pi$-periodic function that represents the conservative part of the chemical dynamics -- that is related to the energy barriers of the washboard potential -- and the second term is the non-equilibrium contribution with $f$ being the chemical driving force related to $\Delta \mu$ through $\Delta \mu = 2 \pi f$. The third term in Eq.~\eqref{eq:potentialU} corresponds to a harmonic potential governing the conformation (elongation) of the enzyme. The rest length $L(\phi)$ (with $L(\phi)$ being a $2\pi$-periodic function) depends on the internal reaction coordinate $\phi$, and is responsible for the reaction-induced conformational changes of the enzyme. The last term of Eq.~\eqref{eq:potentialU} represents the external force $F$ that is applied on one of the sub-units (here sub-unit 1, without loss of generality).

The overdamped dynamics of this system is
\begin{align}
    \dot x_1 = -m_1 \partial_1 U, 
    \\
    \dot x_2 = -m_2 \partial_2 U, 
    \\
    \dot \phi = -m_\phi \partial_\phi U,
\end{align}
where $m_1$, $m_2$, and $m_\phi$ are the corresponding mobilities of the coordinates. To proceed further we assume that sub-unit 2 is rigidly attached to a substrate (see Figs.~\ref{fig:figure1}(a,d)). This implies that $x_2$ is constant (as $m_2 \to 0$), and can thus absorbed in the definition of the rest length $L(\phi)$. Moreover, we assume that the enzyme is relatively stiff, i.e $m_1 k \gg m_\phi \Delta \mu$, and therefore the length relaxes relatively more quickly, as compared to the dynamics of the phase $\phi$. This is equivalent to assuming a strong mechano-chemical coupling of the internal cycle and the spatial conformation. We define $\delta x_1 \equiv x_1 - L(\phi)$, which yields $\delta \dot x_1 = \dot x_1 - L'(\phi) \dot \phi$. This allows us to enslave $x_1$ to the dynamics of $\phi$ \emph{via} the stiff enzyme assumption, which implies that $\delta \dot x_1 \simeq 0$ and the actual enzyme elongation closely trails behind the rest length $L(\phi)$.  This projection reduces the deterministic dynamics of the internal phase $\phi$ to a single, closed equation of motion  given by
\begin{equation}\label{eq:phieqdet}
    \dot \phi = - M(\phi) \partial_\phi V(\phi),
\end{equation}
with a modified effective potential $V(\phi)$ and mobility $M(\phi)$ for the phase given by
\begin{align}\label{eq:enz_pot}
    V(\phi) &= -f\phi   - \tilde{V}(\phi) - F L(\phi), \\
    M(\phi) &= \frac{m_\phi}{1+{m_\phi (\partial_\phi L )^2}/{m_1}}.\label{eq:mobility}
\end{align}

Because enzymatic systems operate at the nano-scale, the thermal fluctuations that help the system overcome the free energy barriers in the potential $V(\phi)$ must be included in a thermodynamically consistent way that respects the fluctuation--dissipation theorem. The Langevin equation for the stochastic dynamics, to be interpreted in the Stratonovich convention, takes the form
\begin{equation}\label{eq:phieq}
    \dot \phi = - M(\phi) \partial_\phi V(\phi) + \frac{k_B T}{2}\partial_\phi M(\phi)+  \sqrt{2k_B T M(\phi)}\,\xi,
\end{equation}
where $k_B$ is the Boltzmann constant, $T$ the temperature, and $\xi$ is a Gaussian white noise with $\langle \xi(t) \rangle =0$ and $\langle\xi(t) \xi(t')\rangle = \delta(t-t')$. The second term is the spurious drift which guarantees that in the absence of non-equilibrium driving $(f \to 0)$ the phase $\phi$ will reach a Boltzmann equilibrium distribution \cite{agudo2021synchronization,chatzittofi2023collective}. It is worth noting that for a free enzyme (not bound to a substrate) a similar derivation can be carried out leading to the same phase dynamics, Eq.~\eqref{eq:phieq}. This coarse-graining procedure has also been recently used to derive a model to study synchronization for coupled enzymes \cite{agudo2021synchronization}.

\begin{figure*}[t]
\centering\includegraphics[width=1.0\linewidth]{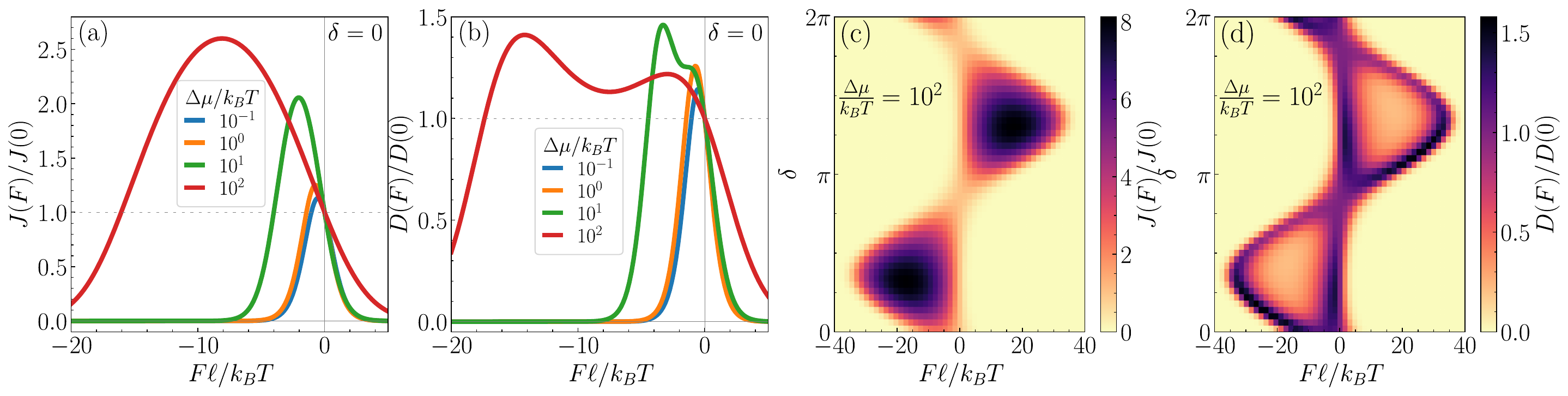}
\caption{(a,b) The internal current (a), and the diffusion of the current (b) for an enzyme that undergoes catalysis-induced conformational changes, as a function of the dimensionless applied force $F\ell/(k_B T)$ for different chemical forces for the case of $\delta = 0$. (c,d) Heat-maps for the current (c) and the diffusion (d) as functions of $\delta$ and the external force for fixed $\Delta \mu/k_B T=10^2$.}
	\label{fig:results_enzyme}
\end{figure*}

For concreteness, we now make a choice for the form of the scalar functions $\tilde V(\phi)$ and $L(\phi)$. To describe the dynamics of a chemical reaction two key energetic scales are needed. One is the chemical driving or Gibbs free energy of the reaction, which is captured by the parameter $\Delta \mu$. The second is the height of the energy barrier $E_\mathrm{ba}$ that must be overcome, which defines a characteristic rate (inverse time scale) of the reaction, namely, the Kramers rate $r_K \propto e^{-E_\mathrm{ba}/k_B T}$. Choosing $\tilde V(\phi) = - v \cos(\phi+\arcsin(f/v))$ leads to the standard form of a washboard potential shown in Fig.~\ref{fig:figure1}(c), given by
\begin{equation}
V_\mathrm{ch}(\phi)= -f\phi - v \cos(\phi+\arcsin(f/v)),
\end{equation}
where the phase shift $\arcsin(f/v)$ is introduced as a convention such that the minima of the chemical potential are located at multiples of $2\pi$. The mapping between $(f,v)$ and $(\Delta \mu, E_\mathrm{ba})$ is given by $\Delta \mu = 2\pi f$ and $E_\mathrm{ba}= [2\sqrt{1-(f/v)^2}-(f/v)\pi]v$ \cite{agudo2021synchronization,chatzittofi2023collective}.

The rest length $L(\phi)$ must be a $2\pi$-periodic function, so that after a full reaction the rest length remains unchanged. At the leading harmonic order, the function $L(\phi)$ can be described by $L = L^{(0)} + \ell \cos(\phi +\delta)$, where $L^{(0)}$ is the average length of the enzyme, $\ell$ is the amplitude of expansion and contraction, and $\delta$ is an arbitrary phase shift. The choice of only the first harmonic implies that, during a catalytic reaction, the enzyme will expand and contract once. For example, if $\delta = 0$ ($=\pi$) then the enzyme is initially in a fully expanded (contracted) state and undergoes one contraction (expansion) during the reaction. As we show on what follows, the phase $\delta$ strongly affects the response to the applied force $F$.

\subsection{Results}

We have managed to reduce the problem to the dynamics of a single degree of freedom. This allows us to use analytical tools from the literature to calculate the average reaction rate $J$ and the diffusion $D$ of the number of reactions at steady-state, defined as
\begin{align}
    J &\equiv \mathrm{lim}_{t\to \infty} \frac{\langle  \phi \rangle_t}{2\pi t}, \\
    D &\equiv \mathrm{lim}_{t\to \infty} \frac{\langle \phi^2 \rangle_t - \langle \phi \rangle_t^2}{(2\pi)^2 \cdot 2t},
\end{align}
where $\langle ... \rangle$ stands for the time average. 

As Eq.~\eqref{eq:mobility} shows, the projection on the slow manifold gives rise to multiplicative noise. Both $J$ and $D$ have been calculated analytically in the case of additive \cite{PhysRevLett.87.010602} and multiplicative \cite{PhysRevE.66.041106} noise. The general expressions for the case with multiplicative noise are
\begin{align}\label{eq:enzymeJ}
J &=  k_B T\,\frac{1-e^{-2\pi f/k_B T}}{\int^{2\pi}_0 dx I_+(x) },\\
D &= m_\phi k_B T \frac{\int^{2\pi}_0 dx \left[I_{+}(x)\right]^2 I_-(x)}{\left[ \int^{2\pi}_0 dx I_+(x)\right]^3},\label{eq:enzymeD}
\end{align}
where we define
\begin{align}
    I_+(x) \equiv \frac{1}{M(x)} \,e^{V(x)/k_B T}\int^x_{x-2\pi} dy \,e^{-V(y)/k_B T},\\
    I_-(x) \equiv e^{-V(x)/k_B T} \int^{x+2\pi}_x dy \,\frac{1}{M(y)} \,e^{V(y)/k_B T}.
\end{align}

In the following, we focus on different values of $\Delta \mu/k_B T$, $F\ell/k_B T$, and $\delta$, while we fix the dimensionless mobility ratio as $m_\phi \ell^2/m_1=1$ and noise strength as $k_B T/E_\mathrm{ba} =1$. Our results for the effect of an external force $F$ on the catalytic dynamics of an enzyme are summarized in Fig.~\ref{fig:results_enzyme}. More specifically, Figs.~\ref{fig:results_enzyme}(a) and (b) display the average reaction rate and diffusion for $\delta = 0$ as functions of the applied force, for several values of the chemical driving $\Delta \mu$. For all choices of $\Delta \mu$, we observe stalling of both the reaction rate and the diffusion at high forces (independently of the sign, i.e.~$|F|\to\infty$). The reaction rate typically shows a maximum at a finite force value, whereas the diffusion typically shows two peaks.

The stalling as $|F|\to \infty $ can be understood as coming from the last term in Eq.~\eqref{eq:enz_pot}, as the energy barrier in the effective potential goes as $F \ell$ that becomes much larger than $k_B T$, which implies that $J(F)\to 0$ and $D(F)\to 0$. The maxima of the reaction rate and the diffusion at finite force can be understood as coming from a delicate interplay between all three terms in Eq.~\eqref{eq:enz_pot}, or more specifically, between $F$, $\delta$, $f$, and $v$. As a particularly striking illustration, let us consider the special case where $\delta = \arcsin(f/v)$ and the elongation $L(\phi)$ is in (anti-) phase with $\tilde{V}(\phi)$. When the external force reaches the critical value $F= - v/\ell$, the second and third terms in Eq.~\eqref{eq:enz_pot} cancel each other, and the internal phase experiences only the constant force $f$. This special case corresponds to the global maximum of $J(F)$ of the heat-map (dark region) in Fig.~\ref{fig:results_enzyme}(c) and local minimum of $D(F)$ (light region) in Fig.~\ref{fig:results_enzyme}(d). On the other hand, when $F= (\pm f - v)/\ell$, the effective potential develops an inflection point which leads to the phenomenon known as giant diffusion \cite{PhysRevLett.87.010602,PhysRevE.65.031104} and explains the two peaks in $D(F)$.

\section{Rotors}

\subsection{Model}

With a similar kind of model as used for an enzyme, we can derive the equations of motion for a rotor (Fig.~\ref{fig:figure1}(b)), representing e.g.~F1-ATPase \cite{junge2015atp,Yasuda1998}. The rotor is described by two coordinates: $\phi$, which represents an internal reaction coordinate, and $\theta$ which is the actual angular state of the rotor \cite{chatzittofi2023topological}. The potential describing their coupling in this case  is
\begin{equation}\label{eq:potentialUrot}
    U_\mathrm{rot}(\theta,\phi) =  \tilde{V}(\phi) - f\phi
    - k_\theta\cos(\phi-n\theta) - \tau \theta ,
\end{equation}
where the first two terms are the same as in Eq.~\eqref{eq:enz_pot} and describe the chemical free energy. The third term is a ``toothed gear'' coupling with strength $k_\theta$ between the angle $\theta$ and the internal phase $\phi$. The integer $n$ describes how many reactions it takes to complete a full turn of the rotor (in the case of ATP-synthase, the motor rotates by $120^\mathrm{o}$ per reaction and therefore $n=3$ \cite{junge2015atp,Yasuda1998}). The last term corresponds to an externally applied torque.

The over-damped deterministic equations of motion are given as
\begin{align}
    \dot \theta = -m_\theta \partial_\theta U_\mathrm{rot},\\
    \dot \phi = -m_\phi \partial_\phi U_\mathrm{rot},
\end{align}
where $m_\theta$ is the angular mobility. For strong coupling, i.e.~$m_\theta k_\theta \gg m_\phi \Delta\mu$, the dynamics of $\theta$ is fast relative to the dynamics of $\phi$, and therefore, enslaved to it, such that $\dot \theta \simeq \dot \phi/n$. After further coarse graining and inclusion of the appropriate noise as above, we find that the stochastic evolution of the internal phase is given by
\begin{equation}\label{eq:rotor}
    \dot \phi = -M_\mathrm{rot}\partial_\phi V_\mathrm{rot}(\phi) + \sqrt{2k_B T M_\mathrm{rot}}\,\xi,
\end{equation}
to be interpreted in the Stratonovich convention, where
\begin{align}
    V_\mathrm{rot}(\phi) &= -\phi(f+\tau/n) - v\cos(\phi+\arcsin(f/v)), \label{eq:potential_rot}\\
    M_\mathrm{rot}&= \frac{m_\phi}{1+{m_\phi }/{(m_\theta n^2)}}.
\end{align}
Notice that there is no spurious drift term in Eq.~\eqref{eq:rotor} since $M_\mathrm{rot}$ is constant and, thus, the thermal noise is additive.

\subsection{Results}

\begin{figure}
\centering\includegraphics[width=1.0\linewidth]{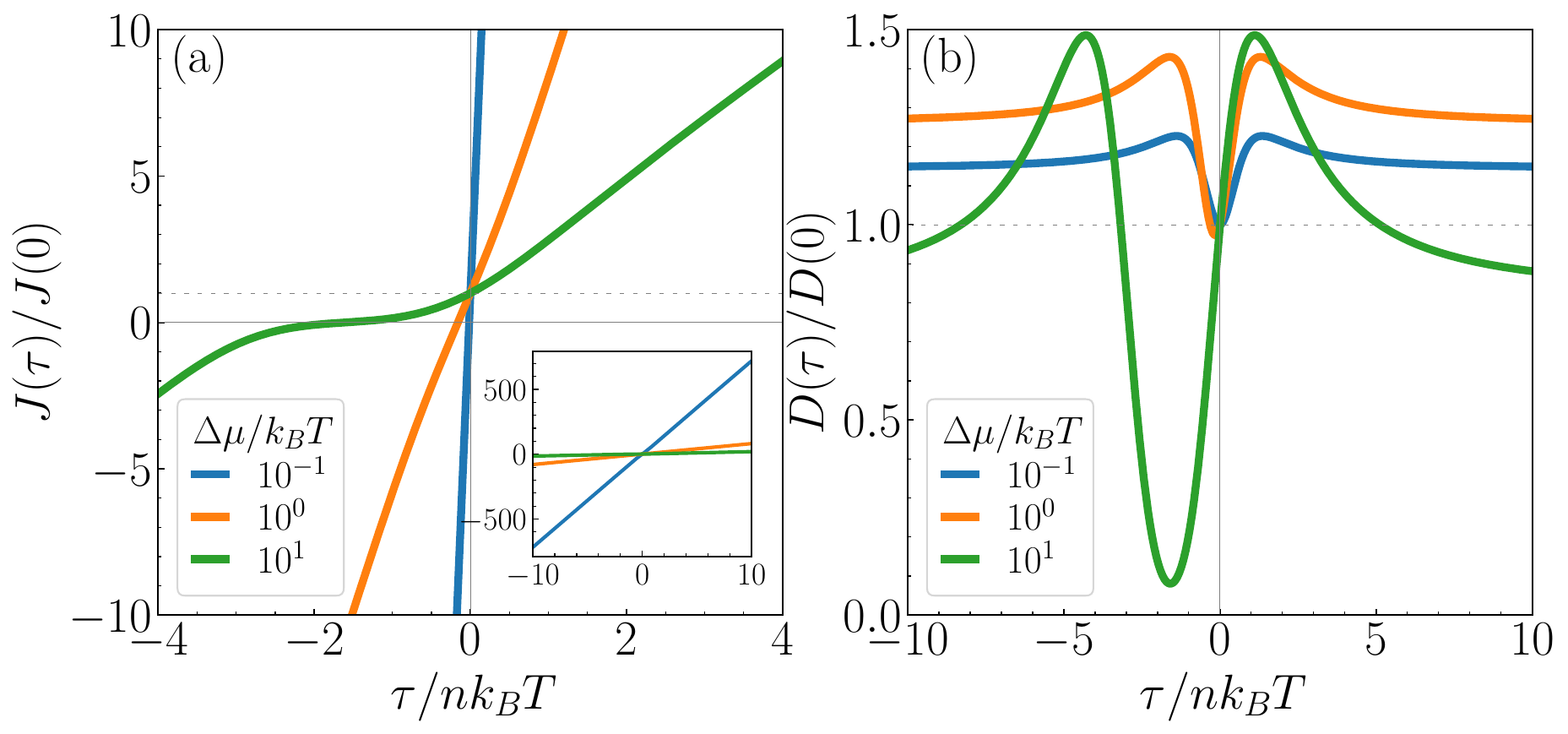}
\caption{(a) The internal current and (b) the diffusion of the current as functions of the external torque for several values of the chemical driving force, for a chemically-driven rotor.}
	\label{fig:results_rotor}
\end{figure}

As Eq.~\eqref{eq:potential_rot} shows, external torques on rotors have a very different effect compared to external forces on enzyme: instead of affecting the energy barriers, the torque $\tau$ now effectively changes the driving force $f$ and thus the chemical affinity $\Delta \mu$ (see Fig.~\ref{fig:figure1}(e)). 
Figure~\ref{fig:results_rotor} summarizes our results for the rotor model. For negative values of the torque $\tau$, which oppose the chemically-driven motion of the rotor, the reaction rate vanishes and switches sign (Fig.~\ref{fig:results_rotor}(a)), implying that the reaction now happens in the reverse direction, e.g.~ATP synthesis instead of hydrolysis.

From Eq.~\eqref{eq:potential_rot}, we see that this reversal occurs at the stall torque $\tau = - n f$ where the effective driving force becomes zero. At high torques, we expect the dynamics to be mostly deterministic, since the energy barriers of the effective potential disappear when $|f +\tau/n| > v$, as shown in Fig.~\ref{fig:results_rotor}(a) (see also the inset) where the current depends linearly on the torque as $\tau \to \pm \infty$. At the stall torque, the diffusion coefficient reaches a minimum since diffusion is lower for a non-driven system (Fig.~\ref{fig:results_rotor}(b)). On either side of this local minimum, we find two peaks in the diffusion, corresponding to giant diffusion occurring at the values of the torque for which the effective potential develops inflection points, $\tau=n(-f \pm v)$. At large values of the torque, the dynamics is effectively reduced to that of a particle under the influence of a constant force, and hence, the diffusion plateaus towards a constant value of $M_\mathrm{rot}k_B T$, as expected (the Einstein relation).

\section{Stochastic micro-swimmers}

\subsection{Model}

In this last example, we consider the Najafi-Golestanian micro-swimmer model \cite{NG2004}, which is made up of three spheres that are linked together by two arms (see Fig.~\ref{fig:figure1}(f)), in its stochastic form \cite{prl2008,golestanian2010,chatzittofi2023entropy}. The motion of the mechanical arms is assumed to be due to an internal non-equilibrium mechanism that dissipates some energy $\Delta \mu$ after a full chemical cycle and drives the non-equilibrium current $J$ as shown in Fig.~\ref{fig:figure1}(f). This coordinated expansion-contraction of the arms ultimately leads to a net displacement of the swimmer after a full cycle, and thus to active swimming. In the absence of a driving affinity $\Delta \mu$, there would be no net current in the chemical cycle, and the micro-swimmer would show no net swimming. Generally speaking, the number of states needed to generate swimming can vary depending on the model, with the minimum number needed to form a cycle being three \cite{golestanian2009stochastic}.

Here, we consider four possible conformations for the swimmer, which are associated with four states in a discrete Markovian stochastic dynamics \cite{prl2008}. The states of the arms are described by the deformations of the left and right arms, $u_\mathrm{left}$ and $u_\mathrm{right}$, respectively, where we assume that the expansions and contractions take place relatively fast. During a conformational transition, one of the two arms expands or contracts by a distance $\ell$. As an example, state $B$ corresponds to $(u_\mathrm{left},u_\mathrm{right})=(0,\ell)$.
The transition from $\alpha \to \beta$ happens with rate $k_{\beta \alpha}$. For a thermodynamically consistent description we must require local detailed balance, given by
\begin{equation}\label{eq:locdetbal}
    \frac{k_{\beta \alpha}}{k_{\alpha \beta}} = \exp(\Delta \sigma_{\beta \alpha}/k_B),
\end{equation}
where $\Delta \sigma_{\beta \alpha}$ is the entropy production during the transition $\alpha \to \beta$ \cite{Seifert2012}. In the absence of external forces on the micro-swimmer, the entropy production is purely due to the chemical and the internal processes. Hence, $T\Delta \sigma_{\beta \alpha} = \Delta \mu_{\beta \alpha}$ and the transition rates satisfy ${k_{0,\beta \alpha}}/{k_{0,\alpha \beta}} = \exp(\Delta \mu_{\beta \alpha}/k_B T)$. The total energy dissipation after a complete cycle in the absence of external forces is then $\sum_{\{\beta \alpha\}} T\Delta \sigma_{\beta \alpha} = \sum_{\{\beta \alpha\}} \Delta \mu_{\beta \alpha} = \Delta \mu$.

\begin{table}[t]
\begin{center}
\caption{Displacement of each sphere during each transition step. For the reverse transitions, $\Delta x_{i,\alpha \beta} = -\Delta x_{i,\beta \alpha}$. 
\label{tab:table}}
\begin{tabular}{cccc} 
     \hline  
     Process & $\Delta x_{1,\beta \alpha}/\ell$ & $\Delta x_{2,\beta \alpha} /\ell$ & $\Delta x_{3,\beta \alpha}/\ell$\\
      \hline 
      $\quad A \longrightarrow B \quad $ & $1-\alpha_L $  & $\quad -\alpha_L \quad$ & $-\alpha_L$\\ 
      $\quad B \longrightarrow C \quad $  & $\quad \alpha_S \quad$ & $\alpha_S$ & $-(1-\alpha_S)$\\
      $\quad C \longrightarrow D \quad $ & $-(1-\alpha_S)$ & $ \quad \alpha_S \quad$ & $\alpha_S$\\
      $\quad D \longrightarrow A \quad $ & $-\alpha_L$ & $\quad -\alpha_L \quad$ & $1-\alpha_L$\\
      \hline 
\end{tabular}
\end{center}
\end{table}

\begin{figure*}[t]
\centering\includegraphics[width=1.0\linewidth]{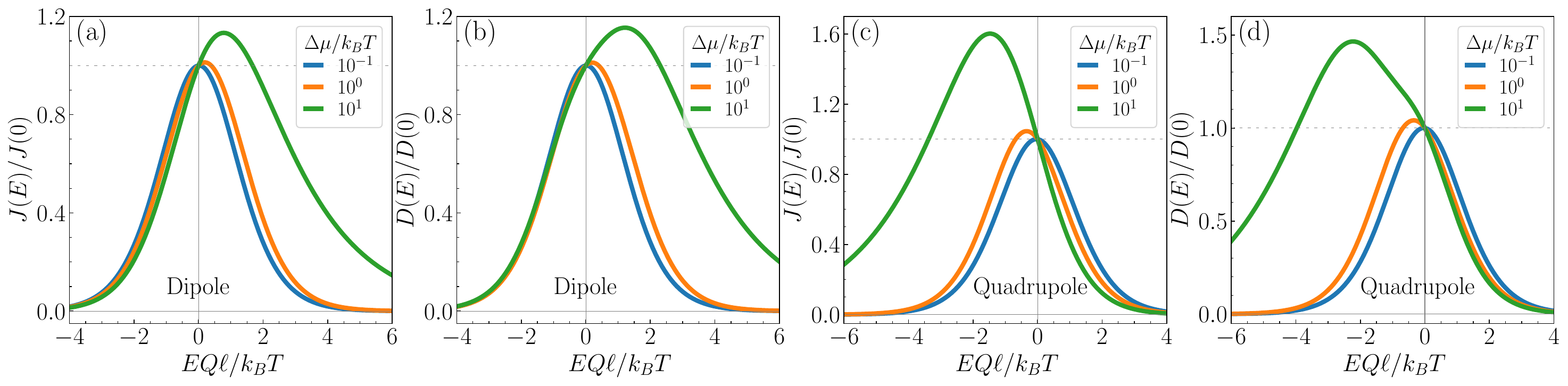}
\caption{(a,c) The internal current and (b,d) the diffusion of the current as functions of an externally applied electric field, for the dipole-model micro-swimmer (a,b) and the quadrupole-model micro-swimmer (c,d); see Fig.~\ref{fig:figure1}(g).}
	\label{fig:results_3ss}
\end{figure*}

To examine the effect of external forces, let us now consider a micro-swimmer with charged components, in the presence of an external electric field, as generalizations of the model presented in Ref. \cite{golestanian2010}. 
We will consider two different classes of micro-swimmers (see Fig.~\ref{fig:figure1}(g)). The first one, namely the ``dipole-model'', has two charged sub-units: the left sphere is negatively charged ($-Q$), the middle one is neutral, and the right sphere is positively charged ($+Q$). The second one, namely the ``quadrupole-model'' has the left and right spheres with negative charge $-Q$, while the central one has positive charge $+2Q$. In both cases, the micro-swimmer is overall neutrally charged. As a consequence, the external field $E$ (parallel to the axis of the micro-swimmer) generates external forces on the charged spheres but does not cause an extra net drift since the net force on the micro-swimmer is zero, i.e.~$\sum_i F_i =0$ where $F_i$ denotes the force on sphere $i$, with $i=1,2,3$. The forces on the spheres, however, modify the work done by the arms during the conformational transitions and thus the dissipation during each step. To calculate this dissipation, we need to know the displacements of each sphere in every transition. These displacements have been calculated in Ref. \cite{chatzittofi2023entropy} through a full hydrodynamic derivation, and are summarized in Table~\ref{tab:table}. The parameters $\alpha_L$ and $\alpha_S$ depend purely on the geometry of the swimmer and satisfy the constraint, $1/3 < \alpha_L < \alpha_S < 1/2$ with the lower bound occurring in the limit of no hydrodynamic interactions (when the spheres remain very far away from each other in all conformations).

The work done by the external field on the spheres in the case of the dipole-model is
\begin{equation}\label{eq:dipole_diss}
    W_{\beta \alpha}=  E Q (\Delta x_{1,\beta \alpha} - \Delta x_{3,\beta \alpha}),
\end{equation}
whereas, for the quadrupole-model, it is
\begin{equation}\label{eq:quadrupole_diss}
W_{\beta \alpha}=  E Q (\Delta x_{1,\beta \alpha} - 2 \Delta x_{2,\beta \alpha} + \Delta x_{3,\beta \alpha}).
\end{equation}
Importantly, because the displacement of every sphere is the same across a full cycle, i.e.~$\sum_{\{\beta \alpha\}} \Delta x_{1,\beta \alpha} = \sum_{\{\beta \alpha\}} \Delta x_{2,\beta \alpha} = \sum_{\{\beta \alpha\}} \Delta x_{3,\beta \alpha}$, we find that there is no net work done by the field after a full cycle, i.e.~$\sum_{\{\beta \alpha\}}W_{\beta \alpha}=0$. Thus, the external field does not contribute to the total affinity driving the cycle, which is still $\Delta \mu$ (see Fig.~\ref{fig:figure1}(g)). However, it still affects the dynamics by modifying the individual transition rates.

To determine the form of the transition rates, we use the expression
\begin{equation}\label{eq:theta}
    k_{\beta \alpha} = k_{0,\beta \alpha} \exp(\theta_{\beta \alpha} W_{\beta \alpha}/k_B),
\end{equation}
where $k_{0,\beta \alpha}$ are the rates in the absence of external forces. Since we must maintain local detailed-balance (Eq.~\eqref{eq:locdetbal}), the parameters $\theta_{\beta \alpha}$ satisfy $\theta_{\beta \alpha} = 1 - \theta_{\alpha \beta}$, and are related to the location of the energy barrier between states $\alpha$ and $\beta$ \cite{chatzittofi2023entropy}.

\subsection{Results}
For the stochastic dynamics of the three-sphere swimmer we consider the Master equation
\begin{align}
    \frac{d P_\alpha}{d t} = \sum_{\beta} (k_{\alpha \beta} P_\beta-k_{\beta \alpha} P_\alpha ).
\end{align}
At steady-state (where we have ${d P^{\rm ss}_\alpha}/{dt} = 0$), the current $J$ is constant, and is given by $J = k_{BA} P^{\rm ss}_A - k_{AB} P^{\rm ss}_B$ (or any other pair of neighbouring states may be used).
An explicit expression for $J$ and the diffusion coefficient $D$ associated with this stochastic current can be found in Refs.~\cite{derrida, koza}. It is worth noting that the self-propulsion speed of the micro-swimmer is proportional to $J$ while its spatial diffusion is closely related to $D$ \cite{chatzittofi2023entropy}.

In the following, we assume $\Delta \mu_{\beta \alpha}=0$ for all transitions except for $\Delta \mu_{DC}= \Delta\mu$. For the displacements, we fix the geometric parameters as $\alpha_L = 2.1/6$ and $\alpha_S = 2.2/6$ (which implies weak hydrodynamic interactions). We further assume $\theta_{\beta \alpha} = 1/2$ for all transitions. The remaining dimensionless parameter are $\Delta \mu/k_B T$ and $E Q \ell /k_B T$. Figure~\ref{fig:results_3ss} summarizes the results for the chemical current $J(E)$ and the corresponding diffusion coefficient $D(E)$ as functions of the applied electric field. Note that we consider $E>0$ to correspond to the electric field pointing towards the left (Fig.~\ref{fig:figure1}(g)), i.e.~against the direction of self-propulsion (whereas for $E<0$ the field points towards the right). The different charge distributions in the two examples give rise to different internal forces during the cycle. Indeed, for the dipole-model swimmer we observe that both $J(E)$ and $D(E)$ peak at positive values of $E$, whereas they peak at negative values for the quadrupole-model swimmer. The enhancement is similar to that observed for our model enzyme above, where the phase shift $\delta$ controlled the optimal force.

In summary, the chemical current, which controls the self-propulsion of the micro-swimmer \cite{prl2008,chatzittofi2023entropy}, exhibits a nonlinear dependence on the applied external field, depending on the charge distribution of the micro-swimmer.

\section{Concluding remarks}

In this work, we have studied three minimal models describing the stochastic, non-equilibrium dynamics of biological and synthetic nano-machines: an enzyme, a rotor, and a micro-swimmer. We focused, in particular, on the nonlinear response of the internal (chemical) dynamics of these nano-machines to externally-applied (spatial) forces. Our results indicate that the average currents and diffusion coefficients describing the reaction dynamics inside these nano-machines exhibit a strongly non-monotonic behaviour in terms of the dependence on the external forces, and typically peak at intermediate values of the applied forces. Interestingly, such non-monotonic behaviour has been reported experimentally for a variety of enzymes, in response to multiple external stimuli \cite{dik2023propelling}. While we developed proof-of-concept models, more elaborate models based on similar principles may be able to quantitatively capture the results of these experiments.

The observed behaviour can be ascribed to several nonlinear mechanisms. For enzymes that undergo conformational changes during catalysis, external forces can effectively reduce the energy barrier in the reaction free energy. In the case of the rotors, an external mechanical torque can effectively increase (or decrease) the driving force of the internal reaction. Lastly, an external electric field can affect the performance of a micro-swimmer even if the latter is neutrally-charged overall, by affecting the energy dissipation and thus the local detailed-balance in individual stochastic transitions between conformations of the micro-swimmer.

The various systems considered here unravel a rich interplay between chemical and mechanical degrees of freedom, which emerges when the mechano-chemical coupling is accounted for in a thermodynamically consistent manner. In the particular case of force-free micro-swimmers, these are subtle but important effects that are lost in models where the swimming is considered to come simply from an ``active force''. We hope that the nonlinear response framework developed here will prove useful in modelling other molecular-scale non-equilibrium systems where a precise formulation of the mechano-chemical coupling is of utmost importance.
 
\acknowledgments
We acknowledge support from the Max Planck School Matter to Life and the MaxSynBio Consortium which are jointly funded by the Federal Ministry of Education and Research (BMBF) of Germany and the Max Planck Society. 

\bibliography{bibfile}

\end{document}